\documentclass{JHEP3}
\usepackage{graphics}
\usepackage{epsfig}
\usepackage{axodraw4j}

\newcommand{\be} {\begin{equation}}
\newcommand{\ee} {\end{equation}}
\newcommand{\bea}{\begin{eqnarray}}
\newcommand{\eea}{\end{eqnarray}}
\newcommand{\bi} {\begin{itemize}}
\newcommand{\ei} {\end{itemize}}
\newcommand{\bc} {\begin{center}}
\newcommand{\ec} {\end{center}}
\newcommand{\bfi}{\begin{figure}}
\newcommand{\efi}{\end{figure}}

\def\Tr{{\rm Tr\,}}
\def\nl{\nonumber \\}

\title{Dark Matter and Pseudo-flat Directions in Weakly Coupled SUSY Breaking Sectors}

\author{
Boaz Keren-Zur${}^1$, Luca Mazzucato${}^2$
  and Yaron Oz${}^1$\\
${}^1$Raymond and Beverly Sackler Faculty of Exact Sciences \\
School of Physics and Astronomy \\
Tel-Aviv University, Ramat-Aviv 69978, Israel\\
${}^2$Simons Center for Geometry and Physics\\
Stony Brook University, Stony Brook, NY 11794-3840, USA\\
E-mails: \email{kerenzu at post.tau.ac.il},
  \email{mazzu at max2.physics.sunysb.edu},
  \email{yaronoz at post.tau.ac.il}
}

\abstract{
We consider candidates for dark matter in models of gauge mediated supersymmetry breaking, in which the supersymmetry breaking sector is weakly coupled and calculable. Such models typically contain classically flat directions, that receive one-loop masses of a few TeV. These pseudo-flat directions provide a new mechanism to account for the cold dark matter relic abundance. We discuss also the possibility of heavy gravitino dark matter in such models.
}


\begin{document}
\section{Introduction}

In theories of gauge mediated supersymmetry breaking \cite{Giudice:1998bp}, the gravitino LSP is light (with mass in the eV to GeV range), due to the low scale of supersymmetry breaking. This leaves a narrow window to realize the gravitino dark matter (the so called sweet spot of ${\cal O}(1)$ GeV \cite{Ibe:2007km}) and therefore poses the problem of identifying alternative possibilities for cold dark matter.

A successful strategy assumes that the supersymmetry breaking sector is strongly coupled and hidden sector baryons or mesons, of mass larger than the TeV scale, provide the cold dark matter \cite{Dimopoulos:1996gy}. On the other hand, recently a lot of attention has been devoted to the study of supersymmetry breaking sectors which are weakly coupled and calculable, following the seminal work of ISS \cite{Intriligator:2006dd}. In this kind of models there is no strong dynamics that can lead to a composite hidden sector hadron responsible for the cold dark matter; hence the necessity to look for alternative weakly coupled candidates for cold dark matter.

In our previous work \cite{Zur:2008zg}, we suggested a new possibility for dark matter in this context. For the specific model analyzed in \cite{Zur:2008zg}, the lifetime of the suggested dark matter candidates turns out to be too short. In this paper we will investigate this possibility in a general setup. The idea is the following. Consider a weakly coupled supersymmetry breaking sector, characterized by a supersymmetric mass scale $M$ and an F-term $F$. We denote the supersymmetry breaking scale by $\Lambda=F/M$. The MSSM soft masses
$$
m_{soft}\sim{\alpha_{SM}\over 4\pi}\Lambda\ ,
$$
are generated at one loop through the messenger interactions. In such weakly coupled models, there are typically pseudo-flat directions (also denoted as pseudo-moduli). These are classically flat directions, that receive mass at one-loop upon supersymmetry breaking 
$$
m^2_L\sim{\alpha_h\over 4\pi}{\Lambda^2} \ ,
$$
where $\alpha_h=h^2/4\pi$ is some small Yukawa interaction in the hidden sector and we denote by $L$ such light fields. Now, these fields may be stable or cosmologically long lived (lifetime $>10^{26}$ sec), with masses typically in the range of a few TeV and they generate a relic density abundance roughly given by
$$
\Omega h^2 \simeq
\left(\frac{10^{-10}GeV^{-2}}{\langle\sigma|v|\rangle}\right)\ .
$$
They may provide therefore the correct dark matter relic abundance $\Omega h^2\sim 0.1$, depending on their (velocity averaged) total annihilation cross section $\langle \sigma |v|\rangle$. The limit on the cross section for them to be not over-abundant is then
$$
\langle\sigma|v|\rangle\gtrsim 10^{-9} - 10^{-10} \,\,{\rm GeV}^{-2} \ .
$$
In this paper, we discuss the annihilation cross sections of the light fields coming from the pseudo-flat directions and show that in some cases they can be viable cold dark matter candidates, namely:
\begin{itemize}
\item If the light fields are charged under the SM gauge group, and their mass is around a TeV, then the standard WIMP mechanism is at work.
\item If they are singlets under the SM gauge group, then they can produce a correct relic abundance in several cases. In the case in which the messenger scale $M$ in the gauge mediation scenario is below
$$
M\lesssim10 N_m\,\,{\rm TeV} \ ,
$$
where $N_m$ is the number of messengers, then the annihilation cross section is into the MSSM Higgs through a messenger loop, assuming that the messenger fields couple to the MSSM Higgs through ${\cal O}(1)$ Yukawa couplings.
\item In the case in which the light fields are charged under a hidden sector gauge boson, the gauge boson mass needs to be lighter than the light field $L$, which is around $1$ TeV, in particular it accomodates the possibility of a GeV gauge boson, recently discussed in \cite{ArkaniHamed:2008qn}.
\end{itemize}

The paper is organized as follows. We will first discuss the possibility of dark matter in weakly coupled hidden sectors in generality in section 2. We will present the mechanism of generating dark matter from the hidden pseudo-flat directions and briefly discuss also the heavy gravitino and pseudo-Goldstone boson relic abundance in such models. In section 3 we will discuss the viability of a heavy gravitino in the specific example of direct gauge mediation of supersymmetry breaking from a metastable hidden sector proposed in \cite{Kitano:2006xg} and analyzed in \cite{Zur:2008zg}.

{\it Note added}: while we were preparing the manuscript, a pre-print appeared \cite{Shih:2009he}, which has some overlap with section 2 of this paper.

\section{Weakly Coupled SUSY Breaking Sectors}

We work in the framework of general gauge mediation \cite{Meade:2008wd}, in which there are two isolated sectors, the MSSM and the dynamical supersymmetry breaking (DSB) sector, which are coupled by the SM gauge interactions and Yukawa couplings to the Higgs.
The DSB sector is characterized by two scales, the mass scale $M$, and the SUSY breaking scale $\sqrt{F}$ which controls the boson-fermion mass splittings. $\sqrt{F}$ is therefore necessarily smaller than $M$ to avoid tachyons. As in the ISS example \cite{Intriligator:2006dd}, the DSB sector is effectively weakly coupled, in the appropriate Seiberg dual description.
In the following it will be convenient to use the soft SUSY breaking scale $\Lambda\equiv\frac{F}{M}$ as a parameterization for the SUSY breaking effects instead of $\sqrt{F}$. We will keep the discussion at a general level.

\subsection{Spectrum and Scales}
\label{spectrumandscales}
The field content of a weakly coupled supersymmetry breaking sector can be generically parameterized according to the following scheme. Assume that we already expanded the classical fields around the supersymmetry breaking vacuum and let us discuss the spectrum, namely their fluctuations around such vacuum, which includes:
\begin{itemize}
\item The Goldstino superfield $X$, whose $F_X$ auxiliary field gets an expectation value and breaks spontaneously supersymmetry.
\item Heavy fields $H_i$ with tree-level superpotential masses of the order $M$. These include the messenger fields, that couple to the F-term and transmit the supersymmetry breaking to the MSSM.
\item
Light fields $L_i$, which are massless at tree-level, and
obtain masses due to SUSY breaking effects. The scalar components $L_i$ of such chiral superfields are usually referred to as {\it pseudo-flat directions} or {\it pseudo-moduli}. Together with their fermionic superpartners $\psi_{L_i}$, they acquire masses proportional to the SUSY breaking scale
\be\label{masslight}
m^2_{L_i}\sim{\alpha_h\over 4\pi}N_m\Lambda^2 \ ,\qquad
m_{\psi_{Li}}\sim {\alpha_h\over 4\pi}N_m\Lambda\epsilon\ ,
\ee
where $\alpha_h=h^2/4\pi$ is a loop factor and $h\sim{\cal O}(1)$ is a hidden sector Yukawa coupling and $N_m$ is the effective number of messengers, defined in \cite{Cheung:2007es} as the ratio between the gaugino and the scalar SUSY breaking scales $\Lambda^2_{1/2}/\Lambda^2_0$. Note that the mass of the fermionic superpartners of the pseudo-flat direction could be further suppressed by a factor $\epsilon$ related to the amount of R-symmetry breaking (in ISS \cite{Intriligator:2006dd} e.g. the accidental R-symmetry sets $\epsilon=0$). We will consider the case of large R-symmetry breaking $\epsilon\sim1$ in the following discussion. The light fields that are either cosmologically long lived (lifetime $\tau>10^{26}s$) or stable (protected by some hidden global symmetry) have some relic abundance, that we will estimate below. These are our main new players.
\item Massless Goldstone bosons (GB), $\phi_{GB}$, and axions, $a$, for broken global symmetries, are expected to obtain a small mass due to explicit
symmetry breaking higher-dimensional operators. These operators will be suppressed by
some higher scale in the theory -- a cutoff of the theory, usually taken to be $M_{GUT}$.
In the case in which they are long lived or stable, we need to make sure that their relic abundance does not overclose the universe.
\item A Gravitino ($\tilde G$), which obtains a mass proportional to the vacuum energy
\bea
m_{\frac{3}{2}}=\frac{F}{\sqrt{3}M_{pl}}\sim M\frac{\Lambda }{\sqrt{3}M_{pl}}\ .
\eea
It might be stable, if R-parity is exact, or long lived, if R-parity is broken. We discuss the scenario in which a heavy gravitino is the dark matter in section \ref{gravitino}.
\item Finally, there might be gauge bosons, carriers of some higgsed dark force, that get a mass proportional to the symmetry breaking VEV times the hidden sector gauge coupling.
We assume that the DSB is weakly interacting therefore the gauge couplings must be small.
The VEVs can be related to several mass scales.
The VEV, $v$, can be determined by tree-level dynamics, in which case one would expect the scale to be of order
$h M$, where $h$ is some factor which controls the tree-level interactions and the dark boson mass is
\bea
m_D=gv\sim gh M
\eea

Another option is an effective FI term which is the result of a mixing between a $U(1)'$ gauge group in the DSB
sector and a gauge group in another sector which obtained a non-zero D-term. Finally, there might be an additional scale responsible for the higgsing of a gauge boson, with mass at or below the light fields mass.
\end{itemize}

Let us discuss the bounds on the hidden sector scales. The standard constraint can be obtained from the LEP bound on the Higgs mass
$10\, {\rm TeV} \lesssim \Lambda \lesssim 100 \,\,{\rm TeV}
$.
Another lower bound can be obtained from the bounds on the neutralino mass.
The soft MSSM gaugino masses arise by integrating out the supersymmetry breaking sector at one loop in the MSSM couplings according to
\be\label{softmssm}
m_{1/2}={\alpha_{SM}\over 4\pi} N_{m}\Lambda \ .
\ee
From the bound on the lightest neutralino $m_{\tilde \chi_1^0}\gtrsim 50$ GeV, we obtain a bound on the soft SUSY breaking scale
\bea\label{twoone}
\Lambda \sim \frac{4\pi m_{\tilde \chi_1^0}}{N_{m}\alpha_{SM}} \gtrsim 80 /{N_{m}} \,\, {\rm TeV}\ ,
\eea
where $N_{m}$ is the number of messengers.

Combined with requirement of no tachyonic scalar messengers $\sqrt{F}\lesssim M$
we get a lower bound on the messenger scale
\bea
M \gtrsim 10\,\,{\rm TeV} \ .
\eea

Gauge mediation models have the advantage that the interactions are flavor blind and do not
generate flavor changing neutral currents (FCNC). However, if the SUSY breaking scale is
high enough, then planck suppressed operators might generate visible flavor effects in $K^0-\overline{K}^0$
mixing or in $\mu\to e\gamma$ transitions. The assumption that the Planck suppressed contributions be at the order of $10^{-3}$ of the gauge mediated effect can be translated to an upper bound on the messenger scale \cite{Giudice:1998bp}
\bea
M\lesssim 10^{12} \,\,{\rm TeV} \ .
\eea

Let us estimate the mass of the light fields parameterizing the pseudo-flat directions. Since their mass is generated upon SUSY breaking, by integrating out at one-loop the messenger fields, the dimensionful parameter $\Lambda$ entering (\ref{masslight}) is the same as the one appearing in the soft MSSM masses (\ref{softmssm}). It follows from (\ref{twoone}) that for ${\cal O}(1)$ hidden Yukawa couplings, the masses of the light fields are
\be\label{massL}
m^2_{L_i}\sim{\alpha_h\over 4\pi}N_{m}\Lambda^2 \gtrsim \,\left({10\over \sqrt{N_m}}{\rm TeV}\right)^2\ ,\qquad
m_{\psi_Li}\sim {\alpha_h\over 4\pi}N_{m}\Lambda\gtrsim 1\,{\rm TeV}\ .
\ee

\subsection{Interactions}
To estimate the relic abundance of the long lived or stable species, we need to know the interactions among the fields we described in the previous section.
We now classify the relevant interactions in a weakly coupled supersymmetry breaking sector, for which we take the Kahler potential to be approximately canonical. As before, we consider the interactions among the fluctuating fields around the supersymmetry breaking vacuum.
\begin{itemize}
\item Tree-level renormalizable F-term interactions

We will consider the following terms in the superpotential
\bea\label{ops}
~ \mu^2X\ , ~h_{ij}XH_i H_j\ ,
~ m_{ij}H_iH_j\ ,~h_{ijk}H_iH_jH_k\ ,~\tilde h_{ijk}L_iH_jH_k\ , ~ \hat h_{ijk}L_iL_jH_k \ ,
\eea
This set of possible terms is dictated by the assumption the the fields $L_i$ are massless at tree-level. Although
Yukawa interactions of the kind $y_{ijk}L_iL_jL_k$ are
non-generic, they would generate some tree-level cross section in the hidden sector, that could be large enough to give the correct relic abundance. Generically the Yukawa's are of ${\cal O}(1)$ and the masses of heavy fields are of order $M$, the hidden sector supersymmetric scale. The linear coupling is proportional to the F-term.

\item Higgs couplings

In weakly coupled models, the simplest possibility that we consider is to require a coupling of the messenger fields (with appropriate SM quantum numbers) to the MSSM Higgs $h_u$ and $h_d$
\be
\lambda_{u,ij} h_u H_i H_j \ ,\qquad \lambda_{d,ij}h_d H_i H_j \ .\label{messhiggs}
\ee
Such couplings will be crucial to obtain the correct annihilation cross sections.

\item Gauge interactions

Assuming that the DSB sector does not introduce additional weakly interacting massless gauge bosons,
we assume that all the additional gauge symmetries are Higgsed.

Particles charged under the MSSM gauge group will also interact via MSSM gauge bosons.

\item Goldstone boson interactions

The Goldstone bosons have a derivative coupling to the conserved current of the broken symmetry.
The interaction will be suppressed by the symmetry breaking VEV.

\end{itemize}

\subsection{Dark Matter and Pseudo-flat Directions}

The abundance of a cold relic is determined mainly by its annihilation cross section into particles in the thermal bath. For particles with $S$-wave annihilation, it is given by \cite{Kolb:1988aj}
\bea
\Omega h^2 &\simeq&
\left(\frac{10^{-10}GeV^{-2}}{\langle\sigma|v|\rangle}\right)\frac{x_f}{g_*^{1/2}}
\eea
where $\langle\sigma|v|\rangle$ is the annihilation cross section, $x_f$ is the mass of the particle divided by the freeze-out temperature (for cold relics it is usually in the range $x_f\sim 20-50$) and $g_*$ is the number
of relativistic degrees of freedom at freeze-out (where the MSSM contributes $g_*\sim200$). Taking into account
the additional degrees of freedom in the DSB sector, we get $\frac{x_f}{g_*^{1/2}}\sim1-10$.
The following discussion is based on dimensional analysis which will be correct up to ${\cal O}(1)-{\cal O}(10)$ parameters, so from now on we arbitrarily assume $x_f=25$, and $\frac{x_f}{g_*^{1/2}}=1$.
Note that for $P$-wave annihilation, the relic density is multiplied by a factor of $2x_f$.

The requirement that dark matter is not over-abundant can now be translated to
\bea
&&\langle\sigma|v|\rangle \gtrsim 10^{-9} - 10^{-10}{\rm GeV}^{-2} \qquad\mathrm{(S-wave)}\label{CScondition}
\eea
Note that while a stable particle with larger cross sections is harmless (although it does not solve the
dark matter problem), smaller cross sections mean over abundance.
We will first present the relic abundance computation of the pseudo-flat directions. We will then discuss the case of dark matter in the form of a heavy gravitino. In the last part we will briefly address the question of the cosmological constraints on the GB's.

We need to distinguish two main frameworks, namely the annihilation into MSSM particles and the annihilation into hidden sector particles. The two cases may be combined, in which case we just pick the largest cross section. The correct relic abundance can be achieved also through non-thermal production, however we will not consider this possibility here.

\subsubsection{Annihilation into MSSM}

Let us evaluate the annihilation cross section of the light fields, namely the pseudo-flat directions that obtain masses at one-loop, to the MSSM. This is very different, depending on whether the light fields are charged or not under the MSSM gauge group.

{\it MSSM-Charged Fields}

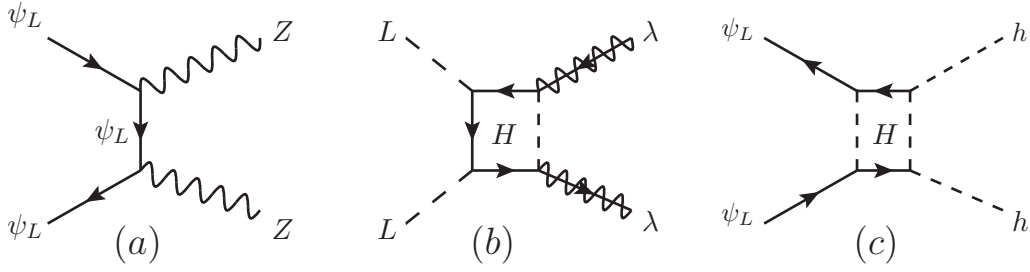
\begin{figure}
\begin{center}
  \begin{picture}(417,82) (29,-18)
    \SetWidth{1.0}
    \Line[arrow,arrowpos=0.5,arrowlength=5,arrowwidth=2,arrowinset=0.2](30,62)(65,42)
    \Line[arrow,arrowpos=0.5,arrowlength=5,arrowwidth=2,arrowinset=0.2](65,42)(65,12)
    \Line[arrow,arrowpos=0.5,arrowlength=5,arrowwidth=2,arrowinset=0.2](65,12)(30,-8)
    \Photon(65,42)(110,62){4}{5}
    \Photon(65,12)(110,-3){4}{5}

    \Line[arrow,arrowpos=0.5,arrowlength=5,arrowwidth=2,arrowinset=0.2](190,42)(190,12)
    \Line[arrow,arrowpos=0.5,arrowlength=5,arrowwidth=2,arrowinset=0.2,flip](190,42)(215,42)
    \Line[arrow,arrowpos=0.5,arrowlength=5,arrowwidth=2,arrowinset=0.2,flip](215,12)(190,12)
    \Photon(215,42)(250,62){4}{5}
    \Photon(215,12)(250,-3){4}{5}
    \Line[arrow,arrowpos=0.5,arrowlength=5,arrowwidth=2,arrowinset=0.2](250,62)(215,42)
    \Line[arrow,arrowpos=0.5,arrowlength=5,arrowwidth=2,arrowinset=0.2](215,12)(250,-3)
    \Line[dash,dashsize=4](215,42)(215,12)
    \Line[dash,dashsize=6](165,62)(190,42)
    \Line[dash,dashsize=6](165,-8)(190,12)

    \Line[arrow,arrowpos=0.5,arrowlength=5,arrowwidth=2,arrowinset=0.2,flip](335,42)(355,42)
    \Line[dash,dashsize=4](355,42)(355,12)
    \Line[dash,dashsize=4](335,42)(335,12)
    \Line[arrow,arrowpos=0.5,arrowlength=5,arrowwidth=2,arrowinset=0.2](335,12)(355,12)
    \Line[dash,dashsize=4](355,42)(390,62)
    \Line[dash,dashsize=4](355,12)(390,-3)
    \Line[arrow,arrowpos=0.5,arrowlength=5,arrowwidth=2,arrowinset=0.2,flip](335,12)(300,-8)
    \Line[arrow,arrowpos=0.5,arrowlength=5,arrowwidth=2,arrowinset=0.2,flip](300,62)(335,42)

    \Text(55,-23)[lb]{\Large{\Black{$(a)$}}}
    \Text(190,-23)[lb]{\Large{\Black{$(b)$}}}
    \Text(335,-23)[lb]{\Large{\Black{$(c)$}}}

    \Text(15,66)[lb]{\normalsize{\Black{$\psi_L$}}}
    \Text(15,-14)[lb]{\normalsize{\Black{$\psi_L$}}}

    \Text(48,22)[lb]{\normalsize{\Black{$\psi_L$}}}

    \Text(115,61)[lb]{\normalsize{\Black{$Z$}}}
        \Text(115,-14)[lb]{\normalsize{\Black{$Z$}}}

    \Text(155,-14)[lb]{\normalsize{\Black{$L$}}}
    \Text(155,61)[lb]{\normalsize{\Black{$L$}}}
    \Text(198,22)[lb]{\normalsize{\Black{$H$}}}
    \Text(255,-11)[lb]{\normalsize{\Black{$\lambda$}}}
    \Text(255,61)[lb]{\normalsize{\Black{$\lambda$}}}

    \Text(395,61)[lb]{\small{\Black{$h$}}}
    \Text(342,22)[lb]{\normalsize{\Black{$H$}}}
    \Text(285,-9)[lb]{\small{\Black{$\psi_L$}}}
    \Text(285,61)[lb]{\small{\Black{$\psi_L$}}}
    \Text(395,-11)[lb]{\normalsize{\Black{$h$}}}
  \end{picture}
\end{center}
\caption{Various annihilation diagrams into MSSM: (a) Direct annihilation of SM charged, electrically neutral light fermion into MSSM gauge bosons - analogue to the WIMP mechanism. (b) Annihilation of light SM singlet scalars into MSSM gauginos via a loop of messengers. (c) Annihilation of light SM singlet fermions into MSSM Higgs via a loop of heavy messengers. }
\label{MSSMAnnDiagram}
\end{figure}

If they are charged, then the leading interaction with the MSSM is at tree-level and the annihilation diagram is as seen in
fig \ref{MSSMAnnDiagram}(a), where the internal line is the light particle. The cross section is given by
\bea
\langle\sigma|v|\rangle\sim\frac{\pi\alpha_{SM}^2}{m_L^2}\sim 10^{-3}\cdot m_L^{-2}
\eea
Therefore a particle of mass $\sim 1$ TeV satisfies the condition (\ref{CScondition}), and we can get a viable dark matter candidate -- a WIMP. Note that larger masses will lead to over-abundance.

{\it Singlet Fields}

If the pseudo-flat directions $L$ are neutral under the MSSM gauge group, the process involves a messenger loop. The reason is that the internal lines in the diagrams \ref{MSSMAnnDiagram}(b) and (c) must
contain at least two heavy fields due to the assumption, below (\ref{ops}), that there are no Yukawa couplings between the light fields. Let us consider the leading effective operators. We have two different possibilities: annihilation into MSSM Higgs $h$, or annihilation into MSSM gauginos. The annihilation into Higgs occurs through the effective operators
\bea
{\alpha_h\alpha_{\lambda} \over M}{\psi_L}\psi_L hh \ ,\qquad{\alpha_h\alpha_{\lambda}} LLhh \ ,\qquad \qquad{\alpha_h\alpha_{\lambda}\over M} LL\psi_h\psi_h \ , \label{higgsannihilation}
\eea
where $\psi_h$ are higgsinos and $\alpha_\lambda=\lambda^2/4\pi$ depends on the messenger couplings to the Higgs (\ref{messhiggs}) and we can generically take the Yukawa couplings $\lambda\sim{\cal O}(1)$. Under the assumption that the DSB sector is calculable, $\alpha_h$ should be a small number $\lesssim1/10$.
The annihilation cross section of light fermions $\psi_L$ into Higgs, through a messenger loop, is given by
\bea
\langle\sigma|v|\rangle \sim {\pi\alpha^2_{h}\alpha_\lambda^2N_m^2 c_G^2 \over M^2} \ ,\label{higgscross}
\eea
where $N_m$ is the number of messengers and $c_G$ is a group theory factor, depending on the representations of the messenger fields. The over-abundance constraint on the cross section (\ref{CScondition}) leads to the following bound on the messenger masses
\be\label{messbound}
M\lesssim 10 N_m\,\, {\rm TeV} \ ,
\ee
and we see that, in the scenario in which the messengers are light enough, the superpartners of the pseudo-flat directions $\psi_L$ can provide the required dark matter relic abundance.
For the light scalars $L$ annihilating into MSSM higgsinos $\psi_h$, the situation is the following. While the last operator in (\ref{higgsannihilation}) provides a cross section of the same order as (\ref{higgscross}), we cannot compute the finite contribution to the cross section coming from the second operator in (\ref{higgsannihilation}), because it is divergent and it requires a UV completion. In any case, the same bound as (\ref{higgscross}) holds.

The annihilation to MSSM  gauginos $\lambda$ occurs through effective operators
\bea
~\frac{\alpha_{SM}\alpha_h}{M}LL  \lambda \lambda\ ,\qquad
~\frac{\alpha_{SM}\alpha_h}{M^2}{\psi}\psi  \lambda \lambda \ .
\eea
The effective operators including the gauge bosons are further suppressed and we did not list them. For the light scalars $L$, the annihilation into gauginos leads to a reasonable cross section
\bea
\langle\sigma|v|\rangle \sim {\pi\alpha^2_{SM}\alpha_h^2N_m^2 c_G^2 \over M^2}\label{scalarcross}\ ,
\eea
which leads to the same bound as (\ref{messbound}). However, the annihilation of their superpartners $\psi_L$ into gauginos is a dimension six operator and the corresponding cross section
\bea\label{fermioncross}
\langle\sigma|v|\rangle \sim {\pi\alpha_{SM}^2\alpha_h^2N_m^2 c_G^2\over  M^{2}}\left(\frac{m_L^2}{M^2}\right) \ ,
\eea
is too small, because of the further suppression factor of $\frac{m_L^2}{M^2}$. Hence, the annihilation into MSSM gauginos alone would lead to overabundance and, in order to have a good relic abundance, we must require the annihilation into the MSSM Higgs through the couplings (\ref{messhiggs}).

\subsubsection{Annihilation inside the Hidden Sector}
Let us evaluate the largest contribution to the cross section coming from the interactions taking place in the hidden sector. Note that we need to keep the messengers from reheating, or this would lead to messenger overabundance \cite{Dimopoulos:1996gy}, so we assume that the reheating temperature is below the messenger scale $T_R<M$. We only need to estimate the cross section for the annihilation of the light fields coming from the pseudo-flat directions into the other light hidden sector fields.

{\it Hidden gauge bosons}

\begin{figure}
\begin{center}
  \begin{picture}(417,82) (29,-18)
    \SetWidth{1.0}
    \Line[arrow,arrowpos=0.5,arrowlength=5,arrowwidth=2,arrowinset=0.2](30,62)(65,42)
    \Line[arrow,arrowpos=0.5,arrowlength=5,arrowwidth=2,arrowinset=0.2](65,42)(65,12)
    \Line[arrow,arrowpos=0.5,arrowlength=5,arrowwidth=2,arrowinset=0.2](65,12)(30,-8)
    \Photon(65,42)(110,62){4}{5}
    \Photon(65,12)(110,-3){4}{5}

    \Line[arrow,arrowpos=0.5,arrowlength=5,arrowwidth=2,arrowinset=0.2](190,42)(190,12)
    \Line[arrow,arrowpos=0.5,arrowlength=5,arrowwidth=2,arrowinset=0.2,flip](190,42)(225,42)
    \Line[arrow,arrowpos=0.5,arrowlength=5,arrowwidth=2,arrowinset=0.2,flip](225,12)(190,12)
    \Line[dash,dashsize=6](250,62)(225,42)
    \Line[dash,dashsize=6](225,12)(250,-3)
    \Line[arrow,arrowpos=0.5,arrowlength=5,arrowwidth=2,arrowinset=0.2](225,12)(225,42)
    \Line[dash,dashsize=6](165,62)(190,42)
    \Line[dash,dashsize=6](165,-8)(190,12)

    \Line[arrow,arrowpos=0.5,arrowlength=5,arrowwidth=2,arrowinset=0.2,flip](335,42)(335,12)
    \Line[dash,dashsize=4](370,62)(335,42)
    \Line[dash,dashsize=4](335,12)(370,-3)
    \Line[arrow,arrowpos=0.5,arrowlength=5,arrowwidth=2,arrowinset=0.2,flip](335,12)(300,-8)
    \Line[arrow,arrowpos=0.5,arrowlength=5,arrowwidth=2,arrowinset=0.2,flip](300,62)(335,42)

    \Text(55,-23)[lb]{\Large{\Black{$(a)$}}}
    \Text(195,-23)[lb]{\Large{\Black{$(b)$}}}
    \Text(330,-23)[lb]{\Large{\Black{$(c)$}}}

    \Text(15,66)[lb]{\normalsize{\Black{$\psi_L$}}}
    \Text(15,-14)[lb]{\normalsize{\Black{$\psi_L$}}}

    \Text(48,22)[lb]{\normalsize{\Black{$\psi_L$}}}

    \Text(203,22)[lb]{\normalsize{\Black{$H$}}}

    \Text(318,22)[lb]{\normalsize{\Black{$\psi_L$}}}

    \Text(115,61)[lb]{\normalsize{\Black{$Z_D$}}}
        \Text(115,-14)[lb]{\normalsize{\Black{$Z_D$}}}

    \Text(155,-14)[lb]{\normalsize{\Black{$L_i$}}}
    \Text(155,61)[lb]{\normalsize{\Black{$L_i$}}}
    \Text(255,-11)[lb]{\normalsize{\Black{$L_j$}}}
    \Text(255,61)[lb]{\normalsize{\Black{$L_j$}}}

    \Text(375,61)[lb]{\small{\Black{$GB$}}}
    \Text(285,-9)[lb]{\small{\Black{$\psi_L$}}}
    \Text(285,61)[lb]{\small{\Black{$\psi_L$}}}
    \Text(375,-11)[lb]{\normalsize{\Black{$GB$}}}
  \end{picture}
\end{center}
\caption{Annihilation diagrams inside the DSB sector:
(a) Direct annihilation of particles into DSB Higgsed gauge bosons.
(b) Annihilation of light singlet scalars into other light scalars through a loop of heavy fields.
(c) Annihilation of light fermions into Goldstone bosons. }
\label{DSBAnnDiagram}
\end{figure}
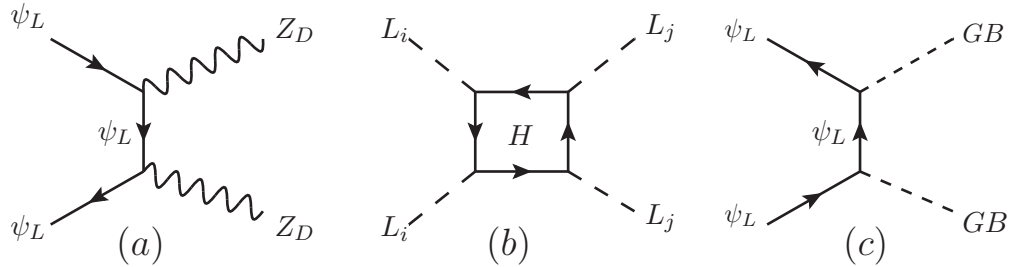

The hidden sector might contain Higgsed gauge bosons with mass $m_D$ and strength $\alpha_D=g_D^2/4\pi$.
If the light fields $L_i$ are charged under this gauge symmetry, they can annihilate into
gauge bosons via tree-level diagrams such as diagram (a) in Fig.\ref{DSBAnnDiagram}. The annihilation cross section will be
\bea
\langle\sigma|v|\rangle\sim\frac{\pi\alpha^2_D}{m_L^2} \ .
\eea
The parameters can easily be set to satisfy the relic abundance condition (\ref{CScondition}), e.g.
\be
\alpha_D\sim 1/100\ ,\qquad m_L\sim 1-10\,\,{\rm TeV} \ .
\ee
Note, however, that the dark gauge boson mass needs to be lower than the light field mass $m_D\lesssim m_L$ for this scenario to be realized. Hence, either the gauge symmetry is higgsed by the heavy fields of mass $M$ at one loop level, or we need to introduce a new scale in the problem that independently sets a light mass for the dark gauge bosons.

{\it Hidden Yukawas}

Let us look at the Yukawa interactions between light and heavy fields. If we allow for several light fields $L_i$, the $LLH$ interaction would produce a tree-level cross section suppressed by the heavy fields, which would be too small. On the other hand, it would also produce one loop annihilation of scalar light fields $LL\to LL$, which is log divergent. Hence, we cannot say anything about this cross section, but one needs to rely on the UV completion of the theory to evaluate it. On the other hand, note that an $LLL$ superpotential interaction, with different species of light fields whose lowest one is stable, would lead to an appropriately large cross section, providing a good dark matter relic abundance. A mechanism of this kind has been recently discussed in \cite{Mardon:2009gw}.\footnote{In \cite{Mardon:2009gw}, the hidden sector is strongly coupled and the dark matter particle are correspondingly heavier, but the cross section is of the same order.}

{\it Hidden GB}

If the light fields are charged under a spontaneously broken global symmetry, they can annihilate
into pseudo-Goldstone bosons. The interaction of the PGB with the Noether current is via
a derivative coupling suppressed by $v$, the symmetry breaking VEV. The annihilation cross sections are given by
\bea
\langle\sigma|v|\rangle \sim\left(\frac{m_{pgb}}{v}\right)^4\frac{1}{m_L^2}~,
\eea
Assuming that $v$ is of the order of the messenger scale $M$,
this is much smaller than the cross sections into MSSM (\ref{scalarcross}) and (\ref{fermioncross}).

\subsection{Heavy gravitino}
\label{gravitino}
In standard gauge mediation scenarios with low energy supersymmetry breaking scale, even if the gravitino mass is larger than ${\cal O}(10)\,{\rm eV}$, its relic abundance can be made negligible. In these cases, the cold dark matter is necessarily a stable particle coming from an additional sector, and in the previous section we provided a new mechanism to generate such cold dark matter.

On the other hand, in the region of parameter space where SUSY is broken at high energy, but still in the framework of gauge mediation, the gravitino might provide a dark matter candidate as well \cite{Ibe:2007km}.
We can easily derive a FCNC upper bound on the gravitino mass in gauge mediation scenario, from the requirement that the gravity contribution to the MSSM soft masses is less than a per mille of the gauge mediation contribution. This leads to the requirement that $m_{3/2}\lesssim20$ GeV. Gravitino dark matter in the GeV range poses a serious cosmological challenge, due to the problems caused by the long-lived NLSP decay with the standard BBN scenario. Nonetheless, there are certain models of gravitino dark matter with mass of ${\cal O}(1)$ GeV, which can be consistently accomodated in gauge mediation and avoid cosmological problems \cite{Ibe:2007km,Lalak:2008bc}. We will see in the next section that this value of the gravitino mass can be consistently accomodated into our specific example of direct gauge mediation.

\subsection{Pseudo-Goldstone bosons}
As discussed in section \ref{spectrumandscales}, the DSB sector might contain Goldstone bosons. Though they are massless
in the limit of exact global symmetry,
their mass will be generated by higher dimensional operators suppressed by the cutoff of the theory, hence we will refer to them as Pseudo-Goldstone bosons (PGB).
The phenomenological implications of such particles require knowledge of
these mass terms, which depend on the details of the UV completion.
On the other hand, the PGB has derivative couplings to the Noether current, suppressed
by the symmetry breaking VEV, which, in the context of our discussion, is of the order of the messenger scale $M$.
This fact will be used in this section in order to obtain some insight on
the phenomenology of the PGB.

If the PGB freeze-out happens after nucleosynthesis, it would contribute to the effective number of relativistic degrees of freedom, $g_*$, which controls the rate of expansion of the universe.
The predictions of BBN are sensitive to this number, and the experimental constraints can accommodate
at most one PGB. If there are more than one, we must make sure that they froze out at higher temperatures $T_f$,
because their contribution to $g_*$ will then be suppressed by $(T_{BBN}/T_f)^4$, where $T_{BBN}$ is around 1 MeV.
Fortunately, an early freeze-out of the Goldstone boson is a very reasonable possibility due to the
fact that the PGB interactions are suppressed by the symmetry breaking VEV, which, as mentioned above, is
about the messenger scale $M$.

If the PGB is stable, we have to find out whether the freeze-out occurred while in the relativistic regime or not, or in other words whether it constitutes a hot or cold relic. Since we do not know the exact mass of the PGB, we will discuss both possibilities.
If the PGB is a hot relic, its contribution to the matter density of the universe depends mainly on its mass
and on the effective number of relativistic degrees of freedom at the moment of freeze-out.
The over-abundance constraint can be translated into a bound on the PGB mass \cite{Kolb:1988aj}:
\bea
m<12.8~eV[g_{*s}(x_f)/g_{eff}]
\eea
The possibility of hot dark matter leads to a tension with structure formation which will not be discussed
here.
The option of the PGB as a cold relic is more problematic, as the annihilation cross section
is suppressed by four powers of the VEV, which will generally lead to overabundance.

We conclude that the early freeze-out required by nucleosynthesis is easily satisfied
for a Goldstone boson in the DSB sector. In order to avoid over abundance
we require the Goldstone boson to be a light hot relic.\footnote{The model discussed in \cite{Ibe:2009dx} contains a cold PGB with mass of around a TeV. The cross section required for the correct relic abundance is obtained by an annihilation process
in which the PGB plays a role of a narrow resonance.}

\section{A Metastable SUSY breaking sector}
\label{koomodel}

In this section we will study a concrete example of a weakly coupled supersymmetry breaking sector, and discuss the possibility of realizing a heavy gravitino of ${\cal O}(1)$ GeV, while simultaneously producing a good MSSM spectrum. This model was suggested in \cite{Kitano:2006xg} as an R-symmetry breaking  deformation of the meta-stable SUSY breaking model of ISS \cite{Intriligator:2006dd}. The phenomenology of this model has been worked out in \cite{Zur:2008zg}, to which we refer the reader for further details.

\subsection{Visible DSB sector and direct mediation}

Consider an $SU(N)$ gauge theory with global $SU(N)\times SU(N_f-N)\times U(1)_B$ and matter content
\bea
\matrix{         & SU(N)       & SU(N)_F     & SU(N_F-N)_F & U(1)_B\cr
  Y              &1            & Adj+1       &1            &  0     \cr
 \hat\Phi             &1            &     1       &Adj+1        &  0     \cr
  Z             &1            & \Box        & \bar\Box    &0       \cr
  \tilde Z      &1            & \bar\Box    & \Box        &0       \cr
  \chi          & \Box        & \Box    &1            &  1     \cr
  \tilde\chi    & \bar\Box    & \bar\Box        &1            & -1     \cr
  \rho          & \Box        &1            & \bar\Box    &  1     \cr
  \tilde\rho    & \bar\Box    &1            & \Box        &-1      \cr
 }
 \eea
The superpotential is given by
\bea
W=h\Tr\big[\tilde\rho\hat\Phi\rho + \tilde\chi Y\chi + \tilde\chi\tilde Z \rho + \tilde\rho Z \chi- m \tilde\chi\chi-\mu\tilde\rho\rho\big]-h\mu^2\Tr\hat \Phi-hm^2\Tr Y +h^2m_z\Tr\tilde Z Z\ ,\nonumber\\ \label{koosuper}
\eea
where the last term is crucial for breaking explicitly the global symmetry and the R-symmetry of the original ISS model.
The theory admits a non-perturbative SUSY vacuum for large values of the fields, the SUSY breaking metastable vacuum of ISS, and several other SUSY breaking vacua, which will not be discussed here. In our previous discussion of the model \cite{Zur:2008zg} we assumed that the universe is in the ISS vacuum,
and chose $N_f=6$, $N=1$, in which case the hidden gauge group is trivial. This construction was used for a direct mediation model, as the gauge group of the standard model was embedded in the $SU(N_F-N)_F$ flavor group. A detailed discussion of the phenomenology led to the following conclusions:
\begin{itemize}
\item
The tension between the longevity of the metastable vacuum and the ratio between gaugino and scalar mass, which appears in the ISS model, is removed
thanks to the introduction of the $R$-symmetry breaking term $m_Z$ which controls the gaugino mass.
By varying this parameter, the model can interpolate between an effective number of messengers ranging from $0<N_m<2$. The price we pay for this solution is that our superpotential is not generic.

\item The pseudo-flat direction $\hat\Phi$, that transforms in the adjoint representation of the GUT $SU(5)$, obtains a mass at the order of 1-10 TeV
due to SUSY breaking effect, and is one example of the light fields in (\ref{massL}). It is therefore potentially detectable in the LHC. Due to its fast decay into
gauge bosons and gravitini, it is not suitable to be a dark matter candidate.

\item Assuming a scenario of cold dark matter, we restricted the model to a low SUSY breaking scale,
which leads to a light gravitino $m_{3/2}<10$ eV. This in turn pushed us to a corner in parameter space where the messengers
are relatively light (near the tachyonic window), and the scalar masses are low. Such a light gravitino is safe from the cosmological point of view, but we need to provide a different dark matter candidate.

\item
Due to the additional  particles in the DSB sector which are charged under the MSSM, the gauge couplings of the MSSM
reach a Landau pole at a scale above the UV cutoff of the theory but below the GUT scale.
A solution to this problem was proposed in terms of a duality cascade \cite{Zur:2008zg} (see also \cite{Csaki:2006wi,Abel:2008tx}).

\end{itemize}

\subsection{Hidden DSB sector}

In the vacuum where $\langle\chi\rangle=\langle\tilde\chi\rangle=hm$, the theory has a $SU(N)_D\times SU(N_f-N)\times U(1)'$ global symmetry.
In the metastable vacuum, the baryon symmetry $U(1)_B$ is broken and there is a messenger $U(1)'$ symmetry with charges $R(\rho)=R(Z)=-R(\tilde \rho)=-R(\tilde Z)=1$.
A messenger parity is also present, namely a $Z_2$ symmetry combined with the charge conjugation in the hidden sector, that acts as follows in four-component spinor notation:
    \bea
    \Psi^p\rightarrow& C(\bar \Psi^p)^T \ ,\nl
    (\rho,Z)\leftrightarrow &(\tilde \rho,\tilde Z)\ , \nl
    \chi\leftrightarrow&    \tilde \chi\ ,
\eea
namely $(\psi_\rho,\psi_Z)\leftrightarrow (\psi_{\tilde \rho},\psi_{\tilde Z})$. Under this symmetry the MSSM gauge vector superfield and the pseudo-flat direction are odd
\bea
 (\Tr\chi_-,\Tr\psi_{\chi_-})\to-(\Tr\chi_-,\Tr\psi_{\chi_-}) \ ,
 \eea
where $\chi_-=\chi-\tilde\chi$,
which allows the decay of $(\chi_-,\Psi_{\chi_-})$ into MSSM particles.
Moreover, at one loop an F-term for $Y$ is generated, so that the Goldstino is a mixture of $\psi_{\hat \Phi}$ and $\psi_Y$ \cite{Essig:2008kz}.

\subsection{Messenger over abundance}

The messengers in the KOO model are a linear combination of the $\rho,\tilde \rho$ and $Z,\tilde Z$ fields. They are the
only fields charged under $U(1)'$,
and are therefore stable.
Stable messengers may be problematic because of their over abundance \cite{Dimopoulos:1996gy}.
One way to avoid this problem is to consider higher order corrections which would break the $U(1)'$ messenger symmetry. In the case of the KOO model, this higher dimension operators are going to be generated at the magnetic cutoff scale or at the GUT scale. Another option is to simply assume that the reheating temperature is below the messenger scale.

\subsection{Heavy gravitino}

The constraints on the SUSY breaking scale which were quoted in the early discussion of the model
($m_{3/2} < 10$ eV in \cite{Zur:2008zg}), were obtained under the assumption that the gravitino plays no role in the cosmology of the model \cite{Viel:2005qj}. In the following we investigate the implications of high scale SUSY breaking in direct mediation models in the specific example of (\ref{koosuper}).\footnote{We would like to thank S. Pokorski for raising this question.} The main additional constraint comes from

\begin{itemize}
\item {Constraints on the parameter space from tachyonic messengers}

In the previous discussion, the combined assumptions of a light gravitino ($\sqrt{F}< 10^5$ GeV) with the requirement of a viable spectrum
($\Lambda\approx\frac{F}{M}>10^5$ GeV) and non-tachyonic messengers ($\frac{F}{M^2}<1$)
pushed us to a very narrow window in parameter space, and required a tuning between the
SUSY breaking scale and the messenger scale. This tension is now reduced when considering high SUSY
breaking scale.

\item {Longevity}

In the KOO model the longevity of the metastable vacuum is controlled by the small parameter $\frac{\mu}{m}$.
This ratio is constrained by the requirement of a fixed value of $\Lambda=\frac{F}{m}\propto\mu\frac{\mu}{m}$.
The option of high SUSY breaking scale (large $\mu$) allows us to take smaller values of this parameter, and enhances the stability of the meta-stable vacuum.\footnote{
Note that this is not a general result. It is relevant only for cases in which the
distance to the next vacuum is proportional to the messenger scale.
In such cases the height of the barrier $V_+$ can be assumed to be proportional to the SUSY breaking scale $\sqrt F=\sqrt{\Lambda M}$, and the bounce action is $
S\sim (\Delta \Phi)^4/V_+\sim \sqrt{ M^{7} /\Lambda}$.
Thus the longevity increases for larger SUSY breaking scales. The effect is reversed for cases in which $\Delta \Phi$ and $M$ are independent: $S\sim (\Delta \Phi)^4/V_+\sim (\Delta \Phi)^4/\sqrt{ \Lambda M}$.
}

\item {Gauge coupling at the messengers scale}

Another constraint comes from the contribution of the DSB sector to the runnning of the MSSM
gauge coupling. Even if we assume that the problem of Landau poles is solved by some
additional mechanism between the cutoff of the magnetic theory and the GUT scale, such as a duality cascade,
the gauge mediation scenario requires the gauge couplings to be perturbative
at the messenger scale for the the soft terms to be calculable.
Using the 1-loop order $\beta$ functions, the requirement $\alpha_s\lesssim 1/4$ at the messenger
scale can be translated to the following bound:
\bea
\alpha_s^{-1}(M_{max})&=&\alpha_s^{-1}(M_Z)+\frac{3}{2\pi}\log\left(\frac{m_{\hat\Phi}}{m_Z}\right)
-\frac{2}{2\pi}\log\left(\frac{M_{max}}{m_{\hat\Phi}}\right)=4\nl
\Rightarrow M_{max}&\approx&10^{11} TeV
\eea
This condition is weaker than the ones obtained from BBN \cite{Giudice:1998bp}
\end{itemize}

\noindent Generally speaking, these considerations must be taken into account in direct mediation models
when there are charged particles in the intermediate scale.

An example for a viable set of parameters with $m_{3/2}\sim1$ Gev is presented in Table \ref{koograv}. The spectrum has been obtained by running the boundary conditions using the modified versions of the SoftSUSY2.0 program \cite{AllanachSOFTSUSY} that was developed in \cite{Zur:2008zg}.

\TABULAR{|c | c | c | c | c | c || c | c | c | c | c|c|}{
\hline
\multicolumn{6}{|c||}{DSB parameters} & \multicolumn{6}{|c|}{MSSM spectrum} \\
\hline
$h$ & $m$  &  $m_z$ & $\mu$ & $\tan\beta$ & $N_c$ &
$m_{3/2}$ & $m[higgs]$ & $m[\chi^0]$ & $m[\tilde g]$ & $m[\tilde t_1] $ &$m[\tilde \tau]$\\ \hline
$2$ & $10^{13}$ & $8 \cdot 10 ^{12}$ &  $2 \cdot 10 ^{9}$ & $5$ & $1$ &
$0.86$& $115$ & $1924$ & $14986$ & $17089$ & $6528$  \\ \hline
}{An example for a set of parameters and the spectrum in a heavy gravitino scenario. All masses are in GeV.}\label{koograv}

We have thus demonstrated that the scenario of heavy gravitino dark matter  is
consistent with the possibility of direct mediation of metastable SUSY breaking.
Nonetheless, it still suffers from the requirement of
fine tuning between the reheating temperature and the SUSY breaking scale.

\acknowledgments We would like to thank N. Itzhaki, S. Nussinov, S. Pokorski and I. Yavin for useful discussions. The work of Y.O. and B.K. is supported in part by the Israeli
Science Foundation center of excellence, by the Deutsch-Israelische
Projektkooperation (DIP), by the US-Israel Binational Science
Foundation (BSF), and by the German-Israeli Foundation (GIF).

\bibliography{susybreak}

\bibliographystyle{JHEP}

\end{document}